\documentclass[twocolumn,
howpacs,preprintnumbers,amsmath,amssymb]{revtex4}

\usepackage{graphicx}
\usepackage{bm}

\def\E^#1{{\buildrel #1 \over\vee}}

\begin{document}

\title{HYDRODYNAMIC EQUATIONS FOR MICROSCOPIC PHASE DENSITIES}%

\author{V.I. GERASIMENKO, V.O. SHTYK$^\dag$, A.G. ZAGORODNY$^\dag$}%

\affiliation{Institute of Mathematics NAS of Ukraine\\
3, Tereshchenkivs'ka str., Kyiv 01601\\
$^\dag$Bogolyubov Institute for Theoretical Physics NAS of Ukraine\\
14-b, Metrolohichna str., Kyiv, 03680}%

\begin{abstract}
The evolution equations for the generalized
microscopic phase densities are introduced.
The evolution equations of average values of microscopic phase densities are derived and
a solution of the initial-value problem of the obtained hydrodynamic type hierarchy is constructed.
\end{abstract}
\maketitle

For the kinetic description of plasmas the method of microscopic phase density is successfully used \cite{Ba05,SZ,ZW09}.
The aim of the present paper is to derive the evolution equations for the generalized
microscopic phase densities and their average values, i.e. the hydrodynamic type equations, in a more consistent way,
and construct solutions of the corresponding initial-value problems.


We consider the system of a non-fixed
(i.e. arbitrary but finite) number of identical particles with unit mass $m=1$ in the space $\mathbb{R}^3$
(\emph{nonequilibrium grand canonical ensemble}).
Every particle is characterized by the phase space coordinates  $x_i\equiv(q_i,p_i)$,
i.e. by a position in the space $q_i\in \mathbb{R}^3$ and a momentum $p_i\in \mathbb{R}^3$.
A description of many-particle systems is formulated in
terms of two sets of objects: by the sequences of observables
$A=(A_0,A_{1}(x_1),\ldots,A_{n}(x_1,\ldots,x_n),\ldots)$ and by the sequences of states
$D=(1,D_{1}(x_1),\ldots,D_{n}(x_1,\ldots,$ $x_n),\ldots)$.
The average values of observables determine a duality between observables
and states. As a consequence, there exist two approaches to the description of the
many-particle system evolution, namely those concerning the evolution of observables or the evolution of states:
\begin{eqnarray}\label{averageDA}
  &&\big\langle A\big\rangle(t)=\\
  &&=\big(1,D(0)\big)^{-1}\sum\limits_{n=0}^{\infty}\frac{1}{n!}
     \int dx_{1}\ldots dx_{n}A_{n}(t)D_{n}(0)=\nonumber\\
  &&=\big(1,D(0)\big)^{-1}\sum\limits_{n=0}^{\infty}\frac{1}{n!}
     \int dx_{1}\ldots dx_{n}A_{n}(0)D_{n}(t),\nonumber
\end{eqnarray}
where
$\big(1,D(0)\big)={\sum\limits}_{n=0}^{\infty}\frac{1}{n!}
\int dx_{1}\ldots dx_{n}D_{n}(0)$ is a normalizing
factor (\emph{grand canonical partition function}).
The sequence $D(t)=(1,D_{1}(t,x_1),\ldots,D_{n}(t,x_1,\ldots,$ $x_n),\ldots)$
of probability densities of the distribution functions  $D_{n}(t)$ is a solution of the initial-value problem of the Liouville
equation. The sequence of observables $A(t)=(A_0,A_{1}(t,x_1),\ldots,$ $A_{n}(t,x_1,\ldots,x_n),\ldots)$
is a solution of the initial-value problem of the Liouville equation for observables.
If $A(0)$ is the sequence of continuous functions and
$D(0)$ is the sequence of integrable functions, then functional (\ref{averageDA}) exists.

An equivalent approach of the description of evolution of many-particle systems,
that enables to describe systems in the thermodynamic
limit, is given by the sequences of $s$-particle (marginal) distribution functions
$F(t)=\big(1,F_{1}(t,x_1),\ldots,F_{s}(t,x_1,\ldots,x_s),\ldots\big)$
and $s$-particle (marginal) observables
$G(t)=\big(G_0,G_{1}(t,x_1),\ldots,$ $G_{s}(t,x_1,\ldots,x_s),\ldots\big)$. The sequence $F(t)$
is a solution of the initial-value problem of the BBGKY hierarchy  \cite{BC},\cite{CGP97},
and $G(t)$ is a solution of the initial-value problem of the dual BBGKY hierarchy \cite{BG}.
In that case, the average values of observables at time moment $t\in \mathbb{R}$ are determined by the functional
\begin{eqnarray}\label{avmarSch}
  &&\big\langle A \big\rangle(t)=
    \sum\limits_{s=0}^{\infty}\frac{1}{s!}
    \int dx_{1}\ldots dx_s G_{s}(0)F_{s}(t)=\\
  &&=\sum\limits_{s=0}^{\infty}\frac{1}{s!}\int dx_{1}\ldots dx_s G_{s}(t)F_{s}(0).\nonumber
\end{eqnarray}
Thus, the sequence of marginal observables $G(t)$ in terms of the sequence $A(t)$ is defined by the formula
\begin{eqnarray}\label{mo}
   &&G_{s}(t,x_1,\ldots,x_s)=\\
   &&=\sum_{n=0}^s\,\frac{(-1)^n}{n!}\sum_{j_1\neq\ldots\neq j_{n}=1}^s
      A_{s-n}\big(t,Y\backslash \{x_{j_1},\ldots,x_{j_{n}}\}\big),\nonumber
\end{eqnarray}
where $Y\equiv(x_1,\ldots,x_s)$,\, $s\geq 1$,
and the sequence $F(t)$ of marginal distribution functions is defined in terms of the sequence $D(t)$ as
\begin{eqnarray}\label{F(D)}
  &&F_{s}(t,x_1,\ldots,x_s)=\\
  &&=\big(1,D(0)\big)^{-1}
     \sum\limits_{n=0}^{\infty}\frac{1}{n!}\int dx_{s+1}\ldots dx_{s+n}D_{s+n}(t).\nonumber
\end{eqnarray}

We remark that, in the case of a system with a fixed number $N$ of particles  (\emph{nonequilibrium canonical  ensemble}) observables
and states are the one-component sequences, respectively, $A^{(N)}=(0,\ldots,0,A_{N},0,\ldots),$
$D^{(N)}=(0,\ldots,0,$ $D_{N},0,\ldots)$. Therefore, the formula for average value (\ref{averageDA})
reduces to the expression
\begin{eqnarray*}
  &&\big\langle A^{(N)}\big\rangle =
     \big(1,D^{(N)}\big)^{-1}\int dx_{1}\ldots dx_{N}A_{N}D_{N},
\end{eqnarray*}
where $\big(1,D^{(N)}\big)=\int dx_{1}\ldots dx_{N} D_{N}$ is a normalizing
factor (\emph{canonical partition function}).

We introduce the observables known as the microscopic phase densities of the system of a non-fixed
number of identical particles.
Let $N(t)\equiv\big(N^{(1)}(t),\ldots, N^{(k)}(t),\ldots\big)$, where
$N^{(k)}(t)=\big(0,\ldots,0,N_{k}^{(k)}(t),\ldots,N^{(k)}_{n}(t),\ldots\big)$,  $k\geq 1$,  is
the sequence of microscopic phase densities of $k$-ary type
\begin{eqnarray}\label{k-arN}
  &&N_{n}^{(k)}(t)\equiv N_{n}^{(k)}(t,\xi_1,\ldots,\xi_k;x_1,\ldots,x_n)=\\
  &&=\sum\limits_{i_1\neq\ldots\neq i_k=1}^{n}\prod\limits_{l=1}^{k}\delta(\xi_{l}-X_{i_l}(t,x_1,\ldots,x_n)),\nonumber
\end{eqnarray}
where $\delta$ is the Dirac $\delta$-function, $\xi_1,\ldots,\xi_k$ are the macroscopic variables
$\xi_i=(v_i,r_i)\in\mathbb{R}^3\times\mathbb{R}^3$. The set of functions
$\big\{X_{i}(t,x_1,\ldots,x_n)\big\}_{i=1}^{n}$, $n\geq k\geq 1$, is a solution
of the Cauchy problem of the Hamilton equations for $n$ particles with the initial data $x_1,\ldots,x_n$ and with the Hamiltonian
$ H_n=\sum\limits_{i=1}^{n}\frac{p_{i}^{2}}{2}+
  \sum\limits_{i<j=1}^{n} \Phi(q_{i}-q_{j}),$
where $\Phi(q_{i}-q_{j})$ is a two-body interaction potential.

For example, if $k=1$, i.e. in the case of an additive-type observable,
we have the microscopic phase density
\begin{eqnarray*}
   &&N_{n}^{(1)}(t,\xi_1;x_1,\ldots,x_n)=\sum\limits_{i=1}^{n}\delta(\xi_{1}-X_{i}(t,x_1,\ldots,x_n)).
\end{eqnarray*}

Microscopic phase densities (\ref{k-arN}) are the solutions of a sequence
of the Cauchy problems of the Liouville equations for observables
\begin{eqnarray}\label{z-K1N}
  &&\frac{\partial}{\partial t}N_{n}^{(k)}(t)=\big(\sum\limits_{i=1}^{n}\langle \,p_i,\frac{\partial}{\partial q_i}\rangle-\\
  &&-\sum\limits_{i\neq j=1}^{n}\langle\frac{\partial}{\partial q_i}\Phi(q_i-q_j),
     \frac{\partial}{\partial p_i}\rangle\big) N_{n}^{(k)}(t),\nonumber
\end{eqnarray}
with the initial data ($1\leq k\leq n$)
\begin{eqnarray}\label{z-K2N}
  && N_{n}^{(k)}(t)|_{t=0}=
   \sum\limits_{i_1\neq\ldots\neq i_k=1}^{n}\,\prod\limits_{l=1}^{k}\delta(\xi_{l}-x_{i_l}),
\end{eqnarray}
where the brackets $\langle \cdot,\cdot\rangle$ denote a scalar product of vectors.

We note that solution (\ref{k-arN}) of Cauchy problem (\ref{z-K1N})-(\ref{z-K2N}) defines
the one-parametric group of operators $\mathbb{R}^1\ni t\mapsto S_n(t)N_{n}(0)$, i.e.
\begin{eqnarray}\label{sn}
   && N_{n}^{(k)}(t,\xi_1,\ldots,\xi_k;x_1,\ldots,x_n)= S_n(t)N_{n}^{(k)}(0),
\end{eqnarray}
where $N_{n}^{(k)}(0)$ is the microscopic phase density (\ref{z-K2N}).

In terms of variables
$\xi_1,\ldots,\xi_k$, the sequence of Liouville equations (\ref{z-K1N}) for microscopic
phase densities (\ref{k-arN}) is represented as the BBGKY
equations set with respect to the arity index $k\geq1$, while it is a sequence of equations with respect to
the index of the number of particles $n\geq k$. Indeed, we have
\begin{eqnarray}\label{qws}
  && \frac{\partial}{\partial t}N_{n}^{(k)}(t)=\big(-\sum\limits_{i=1}^{k}\langle v_i,\frac{\partial}{\partial r_i}\rangle+\\
  &&+\sum\limits_{i\neq j=1}^{k}\langle \frac{\partial}{\partial r_i}\Phi(r_i-r_j),
     \frac{\partial}{\partial v_i}\rangle \big)N_{n}^{(k)}(t)+\nonumber\\
  &&+\sum\limits_{i=1}^{k}\int d\xi_{k+1}\langle\frac{\partial}{\partial r_i}\Phi(r_i-r_{k+1}),
     \frac{\partial}{\partial v_i}\rangle N_{n}^{(k+1)}(t).\nonumber
\end{eqnarray}
If $k=n$, it is the Liouville equation.
Using such a representation of (\ref{z-K1N}), we can directly derive the evolution equations
for average values (\ref{averageDA}) of microscopic phase densities (\ref{k-arN}).
We observe that in the case of a system of a fixed number $N$ of
particles, equations (\ref{qws}) are the hierarchy-type equations set with respect to the index
$k\geq1$, and the Liouville equation with respect to a number of
particles.

The fact that the microscopic phase densities (\ref{k-arN}) are exactly governed by the BBGKY-type hierarchy
of equations (\ref{qws}) is closely connected with the structure of the hierarchy. Indeed, it is known \cite{GP90} that the BBGKY hierarchy
for the states has the explicit solution - the marginal pure state which is the sequence of functions of the phase space variables similar to microscopic phase densities (\ref{k-arN}) as functions with respect to the "macroscopic variables".

We introduce the sequence of marginal observables
$G(t)\equiv\big(G^{(1)}(t),\ldots,$ $ G^{(k)}(t),\ldots\big)$  of $k$-ary type
$G^{(k)}(t)=\big(0,\ldots,0,G_{k}^{(k)}(t),\ldots,G^{(k)}_{s}(t),\ldots\big)$
defined by (\ref{mo}) through microscopic phase densities (\ref{k-arN}).
For example, according to (\ref{mo}) at the initial time moment $t=0$ the sequence of marginal
additive-type microscopic phase densities has the form
$G^{(1)}(0)=\big(0,\ldots,\delta(\xi_{1}-x_1),0,\ldots\big).$
Correspondingly, the sequence of marginal observables
of $k$-ary type microscopic phase densities (\ref{k-arN}) is given as follows:
\begin{eqnarray}\label{ad_G0}
   &&G^{(k)}(0)=\\
   &&=\big(0,\ldots,0,\sum\limits_{i_1\neq\ldots\neq i_k=1}^{k}\prod\limits_{l=1}^{k}\delta(\xi_l-x_{i_l}),0,\ldots\big).\nonumber
\end{eqnarray}

Let $Y\equiv(x_1,\ldots,x_s)$ and
$(x_1,\ldots,\E^{j},\ldots,x_s)\equiv(x_1,\ldots,x_{j-1},x_{j+1},\ldots,x_s)= Y\setminus x_{j}$.
The marginal microscopic phase densities
$G^{(k)}_{s}(t)\equiv G^{(k)}_{s}(t,\xi_1,\ldots,\xi_k;x_1,\ldots,x_s)$  of every $k$-ary type
are governed by the initial-value problem of the dual BBGKY hierarchy \cite{BG}
\begin{eqnarray}\label{dual1}
    &&\frac{\partial}{\partial t}G^{(k)}_{s}(t)=
       \big(\sum\limits_{i=1}^{s}\langle\, p_i,\frac{\partial}{\partial q_i}\rangle-\\
    &&-\sum\limits_{i\neq j=1}^{s}\langle\frac{\partial}{\partial q_i}\Phi(q_i-q_j),
       \frac{\partial}{\partial p_i}\rangle\big) G^{(k)}_{s}(t)-\nonumber\\
    &&-\sum\limits_{i\neq j=1}^{s}\langle\frac{\partial}{\partial q_i}\Phi(q_i-q_j),
       \frac{\partial}{\partial p_i}\rangle G^{(k)}_{s-1}(t,Y\setminus x_{j})\nonumber
\end{eqnarray}
with the initial data
\begin{eqnarray} \label{dual2}
        && G^{(k)}_{s}(t)\mid_{t=0}=G^{(k)}_{s}(0),\quad s\geq k\geq 1.
\end{eqnarray}
As a case in point, we adduce the first equation of hierarchy (\ref{dual1})
\begin{eqnarray*}
    &&\frac{\partial}{\partial t}G^{(k)}_{k}(t)=
       \big(\sum\limits_{i=1}^{k}\langle\, p_i,\frac{\partial}{\partial q_i}\rangle-\\
    &&-\sum\limits_{i\neq j=1}^{k}\langle\frac{\partial}{\partial q_i}\Phi(q_i-q_j),
       \frac{\partial}{\partial p_i}\rangle\big) G^{(k)}_{k}(t).\nonumber
\end{eqnarray*}
For the marginal additive-type microscopic phase density, the first two equations have the form
\begin{eqnarray*}
    &&\frac{\partial}{\partial t}G^{(1)}_{1}(t,\xi_1;x_1)=
       \langle\, p_1,\frac{\partial}{\partial q_1}\rangle G^{(1)}_{1}(t,\xi_1;x_1),\\
    &&\frac{\partial}{\partial t}G^{(1)}_{2}(t,\xi_1;x_1,x_2)=
       \big(\sum\limits_{i=1}^{2}\langle\, p_i,\frac{\partial}{\partial q_i}\rangle-\\
    &&-\sum\limits_{i\neq j=1}^{2}\langle\frac{\partial}{\partial q_i}\Phi(q_i-q_j),
       \frac{\partial}{\partial p_i}\rangle \big) G^{(1)}_{2}(t,\xi_1;x_1,x_2)-\\
     && -\sum\limits_{i\neq j=1}^{2}\langle\frac{\partial}{\partial q_i}\Phi(q_i-q_j),
       \frac{\partial}{\partial p_i}\rangle G^{(1)}_{1}(t,\xi_1;x_{i}).
\end{eqnarray*}

To construct a solution of the dual BBGKY hierarchy (\ref{dual1}), we introduce some preliminaries.
On continuous functions
we introduce the $nth$-order cumulant, $n \geq 1$,
of the groups of operators (\ref{sn})
\begin{eqnarray}\label{cumd}
   &&\mathfrak{A}_{n}(t)\equiv\mathfrak{A}_{n}(t,X)=\\
   &&=\sum\limits_{\mathrm{P}:\, X ={\bigcup}_iX_i}(-1)^{|\mathrm{P}|-1}(|\mathrm{P}|-1)!
     \prod_{X_i\subset \mathrm{P}}S_{|X_i|}(t),\nonumber
\end{eqnarray}
where ${\sum\limits}_\mathrm{P}$
is the sum over all possible partitions $\mathrm{P} $ of the set $X\equiv(x_1,\ldots,x_n)$ into
$|\mathrm{P}|$ nonempty mutually disjoint subsets $ X_i\subset \ X$, the operator $S_{|X_i|}(t)$
is defined by formula (\ref{sn}).

On a continuously differentiable function $g_{1}=g_{1}(x_1)$ the generator of the first-order
cumulant is defined by the operator
\begin{eqnarray*}
    &&\lim\limits_{t\rightarrow 0}\frac{1}{t}\big(\mathfrak{A}_{1}(t,x_1)-I\big) g_{1}=
      \langle\,p_1,\frac{\partial}{\partial q_1}\rangle g_{1}.
\end{eqnarray*}
In the case $n=2$, we have
\begin{eqnarray*}
    &&\lim\limits_{t\rightarrow 0}\frac{1}{t}\,\mathfrak{A}_{2}(t,x_1,x_2) g_{2}=
      -\sum\limits_{i\neq j=1}^{2}\langle\frac{\partial}{\partial q_i}\Phi(q_i-q_j),
      \frac{\partial}{\partial p_i}\rangle g_{2}.
\end{eqnarray*}
If $n>2$, as a consequence of the fact that we consider a system of
particles interacting by a two-body potential the limit
\begin{eqnarray*}
    &&\lim\limits_{t\rightarrow 0}\frac{1}{t}\mathfrak{A}_{n}(t) g_{n}=0.
\end{eqnarray*}
holds.


Let $Y=(x_1,\ldots,x_s),$\, $X=Y\setminus\{x_{j_1},\ldots,x_{j_{s-n}}\}$.
For continuous functions in the capacity of initial data, a solution of Cauchy problem (\ref{dual1})-(\ref{dual2})
is an expansion over particle clusters whose evolution are governed by
the corresponding-order cumulant (semiinvariant) of the
evolution operators of finitely many particles
\begin{eqnarray}\label{rozvG}
    &&G^{(k)}_s(t,Y)=
      \sum\limits_{n=0}^{s}\frac{1}{n!}\sum\limits_{j_1\neq\ldots\neq j_n=1}^{s}\mathfrak{A}_{1+n}\big(t,\\
    &&(Y\setminus X)_1,X\big)G^{(k)}_{s-n}(0,Y\setminus \{x_{j_1},\ldots,x_{j_n}\}), \nonumber
\end{eqnarray}
where the evolution operator
\begin{eqnarray*}
   &&\mathfrak{A}_{1+n}\big(t,(Y\setminus X)_1,X\big)=
     \sum\limits_{\mathrm{P}:\{(Y\setminus X)_1,X\} ={\bigcup\limits}_i X_i}(-1)^{|\mathrm{P}|-1}\\
   &&\times (|\mathrm{P}|-1)!\prod_{X_i\subset \mathrm{P}}S_{|X_i|}(t,X_i)
\end{eqnarray*}
is the $(1+n)th$-order cumulant \cite{GerRS} of groups $S_{|X_i|}(t)$
of operators (\ref{sn}), ${\sum\limits}_\mathrm{P}$
is the sum over all possible partitions $\mathrm{P}$ of the set $\big\{(Y\setminus X)_1,X\big\}$ into
$|\mathrm{P}|$ nonempty mutually disjoint subsets $ X_i\subset \big\{(Y\setminus X)\big)_1,X\big\}$.
The set $(Y\setminus X)_1$ consists of one element of $Y\backslash X$,
i.e. the set $Y\backslash X=\{x_{j_1},\ldots,x_{j_{s-n}}\}$
is a connected subset of the partition $\mathrm{P}$ ($|\mathrm{P}|=1$).

For the additive-type microscopic phase density (\ref{ad_G0}), we derive
\begin{eqnarray}\label{rad}
    &&G^{(1)}_{s}(t,Y)=\mathfrak{A}_s(t,Y)\sum\limits_{j=1}^{s}\delta(\xi_1-x_j).
\end{eqnarray}

Then in terms of variables $\xi_1,\ldots,\xi_k$, the first equation of
hierarchy (\ref{dual1}) for the additive-type microscopic phase density (\ref{rad})
takes the form
\begin{eqnarray*}
   &&\frac{\partial}{\partial t}G^{(1)}_{s}(t,\xi_1;x_1,\ldots,x_s)=\\
   &&=-\langle v_1,\frac{\partial}{\partial r_1}\rangle G^{(1)}_{s}(t,\xi_1;x_1,\ldots,x_s)+\\
 &&+\int d\xi_2\langle\frac{\partial}{\partial r_1}\Phi(r_1-r_2),\frac{\partial}{\partial v_1}\rangle G^{(2)}_{s}(t,\xi_1,\xi_2;x_1,\ldots,x_s),
\end{eqnarray*}
with the initial data
\begin{eqnarray*}
  G^{(1)}_{s}(t,\xi_1;x_1,\ldots,x_s)\mid_{t=0}=\sum\limits_{i=1}^{s}\delta(\xi_1-x_{i})\,\delta_{s,1},
\end{eqnarray*}
where $\delta_{s,1}$ is the Kronecker symbol,\, $s\geq1$.

In a similar manner for the marginal microscopic phase densities of
$k$-ary type $G^{(k)}(t)=\big(0,\ldots,0,G_{k}^{(k)}(t),\ldots,$ $G^{(k)}_{s}(t),\ldots\big)$
we derive
\begin{eqnarray} \label{k_lanc}
 && \frac{\partial}{\partial t}G^{(k)}_{s}(t)=\big(-\sum\limits_{i=1}^{k}\langle v_i,\frac{\partial}{\partial r_i}\rangle+\\
   &&+ \sum\limits_{i\neq j=1}^{k}\langle \frac{\partial}{\partial r_i}\Phi(r_i-r_j),
   \frac{\partial}{\partial v_i}\rangle \big)G^{(k)}_{s}(t)+\nonumber\\
   &&+\sum\limits_{i=1}^{k}\int d\xi_{k+1}\langle\frac{\partial}{\partial r_i}\Phi(r_i-r_{k+1}),
     \frac{\partial}{\partial v_i}\rangle G^{(k+1)}_{s}(t)\nonumber
\end{eqnarray}
with the initial data
\begin{eqnarray} \label{k_lancin}
    &&G^{(k)}_{s}(t)\mid_{t=0}=\sum\limits_{i_1\neq\ldots\neq i_k=1}^{s}
    \prod\limits_{l=1}^{k}\delta(\xi_l-x_{i_l})\delta_{s,k}.
\end{eqnarray}
Here, $1\leq r<s$, and if $k=s$, the marginal microscopic phase density $G^{(s)}_{s}(t)$ is governed by the Liouville equation.

Thus, in terms of variables $\xi_1,\ldots,\xi_k$
the dual BBGKY hierarchy (\ref{k_lanc}) for marginal microscopic phase densities (\ref{rozvG})
is represented as the Bogolyubov equations set with respect
the arity index $k\geq1$,
while evolution equations (\ref{k_lanc}) have a structure of a sequence of equations
with respect to the index of the number of particles $s\geq k$.


We introduce the evolution equations for average values of marginal microscopic phase densities (\ref{rozvG}).
For the $k$-ary type microscopic
phase density, according to (\ref{avmarSch}) and (\ref{rozvG}), from the dual BBGKY hierarchy (\ref{k_lanc}) we derive
the following hydrodynamic type equations for their average values (\ref{avmarSch}):
\begin{eqnarray}\label{Gs1}
    &&\frac{\partial}{\partial t}\langle G^{(k)} \rangle (t)=\big(-\sum\limits_{i=1}^{k}\big\langle v_i,
    \frac{\partial}{\partial r_i}\big\rangle +\\
   &&+ \sum\limits_{i\neq j=1}^{k} \langle \frac{\partial}{\partial r_i}\Phi(r_i-r_j),
    \frac{\partial}{\partial v_i} \rangle\big) \langle G^{(k)} \rangle (t)+\nonumber\\
    &&+\sum\limits_{i=1}^{k}\int d\xi_{k+1} \langle\frac{\partial}{\partial r_i}\Phi(r_i-r_{k+1}),
    \frac{\partial}{\partial v_i} \rangle \langle G^{(k+1)} \rangle (t),\nonumber
\end{eqnarray}
with the initial data
\begin{eqnarray}\label{Gs2}
   && \langle G^{(k)} \rangle (t,\xi_1,\ldots,\xi_k)|_{t=0}=\langle G^{(k)} \rangle(0),\quad k \geq1.
\end{eqnarray}
Due to functional (\ref{avmarSch}), initial data (\ref{Gs2}) are given as the functions
$\langle G^{(k)} \rangle(0,\xi_1,\ldots,\xi_k)=F_{k}(0,\xi_1,\ldots,\xi_k),$
where
$F_s(0,\xi_1,\ldots,\xi_s)$ is the value of initial marginal state at a point $\xi_1,\ldots,\xi_s$.
The hierarchy of equations (18) with respect to the arity index we refer to
as hydrodynamic type equations since it describes the evolution of the average values of the microscopic phase densities.

We note that, according to the definition of functionals (\ref{averageDA}) and (\ref{avmarSch}),
the equality
$\langle G^{(k)} \rangle (t)=\langle N^{(k)} \rangle (t)$ holds in the case of finitely many particles.
In the thermodynamic limit, the value $\langle N^{(k)} \rangle (t)$ tends to $\langle G^{(k)} \rangle (t)$,
i.e. to the solution of Cauchy problem (\ref{Gs1})-(\ref{Gs2}).

A solution of Cauchy problem (\ref{Gs1})-(\ref{Gs2}) is defined by the expansion
over the arity index of the microscopic phase density, whose evolution is governed by
the corresponding-order cumulant of the
evolution operators similar to (\ref{sn}), namely
\begin{eqnarray}\label{RozvG}
&&\langle G^{(k)} \rangle (t,\xi_1,\ldots,\xi_k)=
   \sum\limits_{n=0}^{\infty}\frac{1}{n!}\int d\xi_{k+1}\ldots \\
   &&\ldots d\xi_{k+n}\mathfrak{A}_{1+n}(-t, Y_{1},\xi_{k+1},\ldots,\xi_{k+n})\langle G^{(k+n)}\rangle (0),\nonumber
\end{eqnarray}
where $\langle G^{(k+n)}\rangle (0)$ are initial data (\ref{k_lancin}) and
$\mathfrak{A}_{1+n}(-t,Y_{1},\xi_{k+1},\ldots,\xi_{k+n})$
is the $(1+n)th$-order cumulant of groups of evolution operators similar to (\ref{sn}).
For integrable functions $\langle G^{(k)}\rangle (0)$ series (\ref{RozvG}) converges for small densities.
We note that, if we apply the Duhamel formula to cumulants of groups of operators similar to (\ref{sn}), solution expansion (\ref{RozvG}) reduces
to the iteration series of the hydrodynamic-type hierarchy (\ref{Gs1}).

In summary, the evolution of the generalized marginal
microscopic phase densities (\ref{ad_G0}) is described by the initial-value problem of the dual BBGKY hierarchy (\ref{dual1}).
Their average values (\ref{avmarSch}) are governed by the initial-value problem of hierarchy (\ref{Gs1}) of the hydrodynamic-type equations
which has the structure of the BBGKY hierarchy with respect to the arity index of the generalized
microscopic phase densities. The solutions of the Cauchy problems of such hierarchies (\ref{dual1}) and
(\ref{Gs1}) are represented by the expansions (\ref{rozvG}) and
(\ref{RozvG}) correspondingly.


\end{document}